\RequirePackage{fix-cm}
\documentclass[smallextended]{svjour3}       
\smartqed  
\usepackage{graphicx}
\usepackage{fullpage,authblk}
\usepackage{amssymb,amsfonts,amsmath,enumerate}
\usepackage[bookmarks, colorlinks=true, plainpages = false, citecolor = green, urlcolor = blue, filecolor = blue]{hyperref}
\usepackage{graphicx,subfigure,bm,latexsym}
\usepackage{multirow}

\begin{document}

\title{BiFold visualization of bipartite datasets\thanks{This work was funded in part by the Simons Foundation Grant No. 318812.}
}


\author{Yazhen Jiang   \and
        Joseph D. Skufca \and 
        Jie Sun
}


\institute{Yazhen Jiang \at
              Department of Mathematics, Clarkson University, Potsdam, NY 13699, USA\\
           \and
           Joseph D. Skufca \at
           Department of Mathematics, Clarkson University, Potsdam, NY 13699, USA\\
                 \email{jskufca@clarkson.edu}           
            \and
           Jie Sun \textcolor{black}{(\emph{Corresponding Author})} \at
           Department of Mathematics, Clarkson University, Potsdam, NY 13699, USA\\
           \textcolor{black}{Clarkson Center for Complex Systems Science ($C^3S^2$)}, Potsdam, NY 13699, USA\\
           Department of Physics, Clarkson University, Potsdam, NY 13699, USA\\
           Department of Computer Science, Clarkson University, Potsdam, NY 13699, USA\\
              \email{sunj@clarkson.edu}           
              }

\date{Received: date / Accepted: date}

\maketitle

\begin{abstract}
The emerging domain of data-enabled science necessitates development of algorithms and tools for knowledge discovery. Human interaction with data through well-constructed graphical representation can take special advantage of our visual ability to identify patterns.  We develop a data visualization framework, called BiFold, for exploratory analysis of bipartite datasets that describe binary relationships between groups of objects.  
Typical data examples would include voting records, organizational memberships, and pairwise associations, or other binary datasets. 
BiFold provides a low dimensional embedding of data that represents similarity by visual nearness, analogous  to Multidimensional Scaling (MDS). The unique and new feature of BiFold is its ability to simultaneously capture both within-group and between-group relationships among objects, enhancing  knowledge discovery. We benchmark BiFold using the {\it Southern Women Dataset}, where social groups are now visually evident. We construct BiFold plots for two US voting datasets: For the presidential election outcomes since 1976, BiFold illustrates the evolving geopolitical structures that underlie these election results.  For Senate congressional voting, BiFold identifies a partisan coordinate, separating senators into two parties while simultaneously visualizing a bipartisan-coalition coordinate which captures  the ultimate fate of the bills (pass/fail). Finally, we consider a global cuisine dataset of the association between recipes and food ingredients. BiFold allows us to visually compare and contrast cuisines while also allowing identification of signature ingredients of individual cuisines.
\end{abstract}

\section{Introduction}
\label{intro}
Despite the dominance of automated algorithms for data mining and knowledge discovery,
it has been increasingly recognized that human perception can play an essential and often favorable role in exploring patterns and developing insights~\cite{Fekete2008}. For instance, the Hertzsprung Russell diagram of stellar luminoscity versus temperature provides a classic example of a data analysis problem easily tackled by a person but remains a challenge for automated methods~\cite{Spence1993}. Typically, the utilization of human cognition in exploratory data analysis relies on proper representation and visualization of the data in a low-dimensional embedding space~\cite{Fayyad2001,Gastner2004,Sims2005,Chen2009,Nishikawa2011,Shekhar2014}.

The standard concept of a ``dataset'' is a tabular array, where each row corresponds to an object in the dataset and every column corresponds to a variable (or factor) measured on each object.   A natural question about such a dataset is ``how are objects like (or unlike) other objects and are there relevant relationships among collections of objects?'' 
Multidimensional scaling (MDS) refers to a family of techniques that address these questions by visualizing the {\it objects} as a set of {\it points} embedded in a low-dimensional (typically 2-D or 3-D) geometric space, with the goal of representing the dissimilarities between  objects by the distances between the corresponding  points in the embedded space~\cite{CoxBook,BorgBook}. The generality of MDS approaches makes them suitable for a broad range of practical problems, as demonstrated in many classical examples~\cite{CoxBook,BorgBook} as well as in several recent scientific breakthroughs: mapping of brainwide neural behavior~\cite{Vogelstein2014}, discovery of sex-specific and species-specific perceptual spaces among different biological species~\cite{Engeszer2008}, and analysis of biogeographic differentiation between geographical regions~\cite{Carmeno2009}.
On a more fundamental level, several recent developments focused on generalizing different measures of ``distance" in the MDS formulation to allow for embedding from and/or onto general nonlinear manifolds~\cite{Bronstein2006,Shieh2011,Aflaloa2013}.

Frequently, we encounter dataset which encodes a {\it binary relation} between two sets (or ``classes") of objects, with elements of one set corresponding to the rows, elements of the other corresponding to the columns, and the data entries (``1'' or ``0'') indicating whether or not there is a  relationship between the associated row and column.
Common examples include  {\it politicians} and {\it bills} they supported, or {\it movie-goers} and the {\it movies} that they attend, or {\it students} and the {\it courses} in which they enroll.  Such examples can be regarded as decision-makers and choices, while we note that similar datasets arise in many contexts that are often described by bipartite graphs, such as the association between genes and diseases~\cite{Bauer-Mehren2011}, relation between chemical reactants and  reactions~\cite{Craciun2006}.

Knowledge discovery on binary relation datasets can benefit from a visualization of both decision-makers and choices in a {\it common} embedding space, where (simultaneously)
\begin{enumerate}[(1)]
\item ``Similar'' objects (whether decision-makers or choices) ought to be ``nearby'' in the visualization;
\item Decision-makers should be positioned ``close'' to their preferred choices.
\end{enumerate}
The BiFold method developed here relates to a set of ordination methods that attempts to resolve various aspects of this challenge.  From a classical perspective, one should build that framework upon three primary choices, where we would point the reader to \cite{CoxBook,BorgBook} and the references therein for details of common methods:
{\it Biplot}~\cite{Gabriel1971} aims to satisfy requirement (1), with points (typically referred to as ``samples'') representing one set of objects and coordinate axis (often referred to as ``variables" and plotted as position vectors) describing the other set; {\it Unfolding}~\cite{Bennett1960} considers only between-class distances, and therefore focuses only on requirement (2);  {\it Correspondence analysis}~\cite{Hotelling1933,Richardson1933,Hirschfeld1935} focused on contingency table data rather than general binary relation data.
The BiFold method developed herein merges the respective goals of \underline{Bi}plot and Un\underline{fold}ing  methods, satisfying both requirements (1) and (2), and it is this connection that motivates  the name.  

In addition to the classical ordination methods described above, we note that BiFold has similar goals to nonlinear and generalized biplot methods described in~\cite{Gower1988,Gower1992}.  In particular,  the generalized biplot addresses categorical variables (of which dichotomous variables are a subset) with consideration of both requirements (1) and (2) in developing the ordination.  Their approach is to ordinate each entry in the dataset, such that each level of a categorical variable is separately visualized.  For binary variables, that approach would require representation of \textit{one} of the classes of objects by two sets of ordination coordinates, one to represent the ``1'' and the other to represent the ``0'' in the data.
Our \textcolor{black}{original} contribution here is two-fold: (i) Our treatment is completely symmetric with respect to the classes, with neither being treated as the ``variables.''  The resulting ordination is identical, even if we transpose our dataset.  Consequently, each object, regardless of class, is assigned only one coordinate. (ii) We consider an ordination scheme that accounts for the difference in information quality of cross-group and within-group distances, as well as the difference in information content across groups of different size, as specialized to the binary data framework.  The ordination approach is more naturally able to account for the difference in information content arising from the non-square data matrices, missing data, and differences in interpretation of matches for categorical variables  \cite{Gower1992}. 

We begin with an introductory example of BiFold below, leaving the details of the approach and more examples to the later sections.

\begin{figure}[htbp]
\centering
\includegraphics[width=6in]{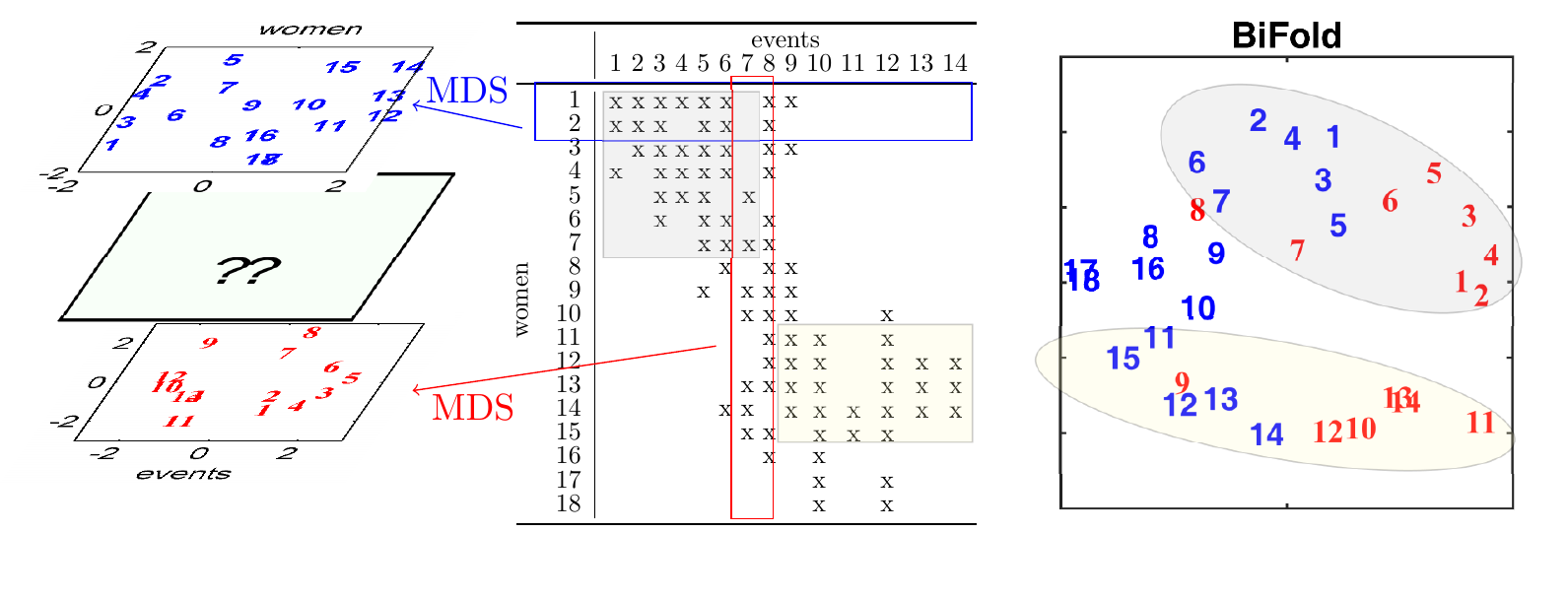}
\caption{{\bf BiFold as a joint visualization.} (Middle panel) The classic {\it Southern Women Dataset}~\cite{SWdata}, where Davis has arranged the table to highlight the social group structures. (Left panel) Independent MDS representations can be created for the women (using inter-row distance) or for the events (using inter-column distance), but those plots cannot be directly merged because the coordinate axes are not same.  (Right panel) The BiFold representation gives coordinate representation against a common basis.  {\it Note} that the significant ``clusters'' observable in the sorted data matrix (colored rectangles) are matched to spatially clustered sets (colored ellipses) in the BiFold plot.  \label{Fig_SW}}
\end{figure}

\medskip\noindent{\bf An Introductory Example - BiFold Plot of the Southern Women Dataset.}
Consider the {\it Southern Women} dataset, collected in the 1930s in a small town in the southern United States. The data records the participation of $18$ ladies (Southern Women) in $14$ social events \cite{SWdata} and can be represented by  matrix $B=[b_{ij}]_{18\times14}$, where $b_{ij}=1$ indicates that woman $i$ attended event $j$, and $b_{ij}=0$ otherwise (see Fig.~\ref{Fig_SW} middle panel as well as {\it Materials and Methods}). Due to its relatively small size and simple structure, the dataset serves as a popular benchmark for techniques that consider social stratification, group formation, and other social structure questions~\cite{Freeman2003}.

One way to visualize the Southern Women dataset is to use MDS to place the 18 women at suitable 2-D locations, where distance between embedded coordinates reflects the degree to which the women attended similar events, as in Fig \ref{Fig_SW}--top left. 
In this case, we are treating the {\it women} as the entities to be plotted, while the {\it events} are regarded as {\it factors} that characterize each individual. Alternatively, we can treat the events as entities and the women as factors, allowing us to obtain an MDS configuration of events, as shown in Fig.~\ref{Fig_SW}--bottom left.
The goal is to ``overlay'' the two embeddings to not only capture the within-class relationships (woman to woman, event to event) but also the cross-class relationship of woman to event. 
BiFold produces such a joint visualization (Fig.~\ref{Fig_SW}--right panel) in which social group structure \cite{Freeman2003,Field2006}  is easily identified through {\it proximity}: (a) nearby women attended similar events, (b) nearby events were attended by similar groups of women, (c) nearby woman-event pairs indicate that the woman likely attended that event.

\section{Results}
\subsection{The BiFold Approach\label{sAlgorithm}}
BiFold provides a procedural framework to produce a low-dimensional embedding from a binary data matrix. 
First, we create a {\it joint} dissimilarity matrix that appropriately fuses information from both within-class and cross-class relations. Secondly, we construct a weighting matrix to reflect the relative uncertainty associated with the dissimilarities. Finally, we minimize a weighted stress function to obtain a {\it BiFold embedding}, coordinates in $\mathbb{R}^d$ for each row and each column of the data matrix.  In this section, we describe and explain this {\it framework}, leaving the detailed specification of the algorithms and parameters to {\it Materials and Methods}.  

Given a binary relation between two types (classes) of objects, encoded as matrix $B=[b_{ij}]_{m\times n}$, where 
\begin{equation}
b_{ij}=
\begin{cases}
1,\mbox{~if object $i$ of type 1 relates to object $j$ of type 2;}\\
0,\mbox{~otherwise}.
\end{cases}
\end{equation}
Such data equivalently encodes a bipartite graph, where an edge in the graph corresponds to a binary relationship, and the matrix $B$ is the {\it biadjacency matrix} of the graph~\cite{BeinekeBook}.

Focusing on objects represented by the rows of $B$, we quantify, using some appropriate measure, the {\it dissimilarity} between row $i$ and row $j,$   denoted $\delta^{(x)}_{ij},$ producing matrix $\Delta^{(x)}=[\delta^{(x)}_{ij}]_{m\times m}.$
Likewise, we generate dissimilarity matrix $\Delta^{(y)}=[\delta^{(y)}_{ij}]_{n\times n}$ by comparing the columns of $B.$ 
Finally, the dissimilarity between row object $i$  and column object $j$  is defined by a monotonic transformation of the entries in matrix $B$, which yields a cross-class dissimilarity matrix $\Delta^{(xy)}=[\delta^{(xy)}_{ij}]_{m\times n}$. 
A binary relation dataset typically falls into one of the two categories: (1) choice data, for which each data entry (whether  ``$0$" or a ``$1$")  reflects an active decision (either a positive or negative relation); and (2) association data, for which the ``$0$"s  indicate only an absence of a relation, and are usually much less informative than the ``$1$"s. 
For data from each of these categories, we have developed some sensible dissimilarity measures (see {\it Materials and Methods}).

Given within-class dissimilarity matrices $\Delta^{(x)}$ and $\Delta^{(y)}$ together with the cross-class dissimilarity matrix $\Delta^{(xy)}$, we form a {\it joint dissimilarity matrix} of size $(m+n)\times(m+n)$, as:
\begin{equation}\label{eqDefDist}
\Delta = \begin{bmatrix}
      \alpha_x\Delta^{(x)} & \alpha_{xy}\Delta^{(xy)}+\beta\bm{1}       \\[0.2em]
      \alpha_{xy}\Delta^{(yx)}+\beta\bm{1} &   \alpha_y\Delta^{(y)} \\
     \end{bmatrix},
\end{equation}
where $\bm{1}$ is a matrix of $1$s, and $\Delta^{(yx)}$ is the matrix transpose of $\Delta^{(xy)}$. 
The joint dissimilarity matrix contains a few tunable parameters: $\alpha_x$, $\alpha_y$, and $\alpha_{xy}$ control the relative scale of the within-class and cross-class distances in the embedded space, while  $\beta$ allows for explicit translation \textcolor{black}{(i.e., shifting)} of the row objects away from the column objects.  In the examples of this paper, we use $\beta = 0$ (no translation). Note, however, that for some datasets the visualization might be improved by a translation (realized by a nonzero $\beta$ value) \textcolor{black}{as determined by the end-user}.

Dissimilarities are generated from {\it data} and should be viewed as a {\it measurement with uncertainty}.
To capture such uncertainty, we associate a ``weight'' to each dissimilarity following the principle that the weight should reflect the information content (or reliability).  We denote the corresponding  joint weighting matrix as
\begin{equation}\label{eq:weightmatrix}
W = \begin{bmatrix}
     W^{(x)} & W^{(xy)}  \\[0.2em]
      W^{(yx)} &  W^{(y)} \\
     \end{bmatrix}.
\end{equation}

Once the joint dissimilarity and weighting matrices are specified,  a $d$-dimensional BiFold embedding yields coordinates $X=(\bm{x}_1,\bm{x}_2,\dots,\bm{x}_m)$ and $Y=(\bm{y}_1,\bm{y}_2,\dots,\bm{y}_n)$ to denote the sets of points corresponding row objects and column objects.  Denote the full coordinate set as $Z=(X,Y).$  Such an embedding is computed by  minimization of the multivariate stress function
$S:\mathbb{R}^{d\times(m+n)}\rightarrow\mathbb{R}$, defined as
\begin{equation} \label{eq:stress}
S(\bm{z}_1,\bm{z}_2,\dots,\bm{z}_{m+n})=\sum_{k,\ell=1}^{m+n}w_{k\ell}(\|\bm{z}_k-\bm{z}_\ell\|-\delta_{k\ell})^2.
\end{equation}
Here $\|\cdot\|$ denotes a metric distance in the common embedding space. \textcolor{black}{The stress function $S$, which is given by a weighted sum of the discrepancies between the embedded distances and the dissimilarities, is a standard type of loss function frequently used in MDS~\cite{CoxBook,BorgBook}.}

In the following sections, we illustrate BiFold via three additional binary datasets: US presidential election results,  US senate voting records from the 112th Congress, and a food-recipe relational dataset from five major global cuisines.

\subsection{Examples: Voting Datasets}
A common type of binary relation comes from voting data: for each item to be voted on, a voter either votes ``for'' or  ``against,'' with (perhaps) the ability to abstain.  We consider two such examples: US presidential election results for past ten elections, for which the ``voters" are the individual US states and the ``items" are the winning presidents in each election.  We also examine senate congressional roll call votes,  with US senators as voters and the items are senate bills. 

\subsubsection{Presidential election by states.}
\begin{figure}[htbp]
\centering
\includegraphics[width=6in]{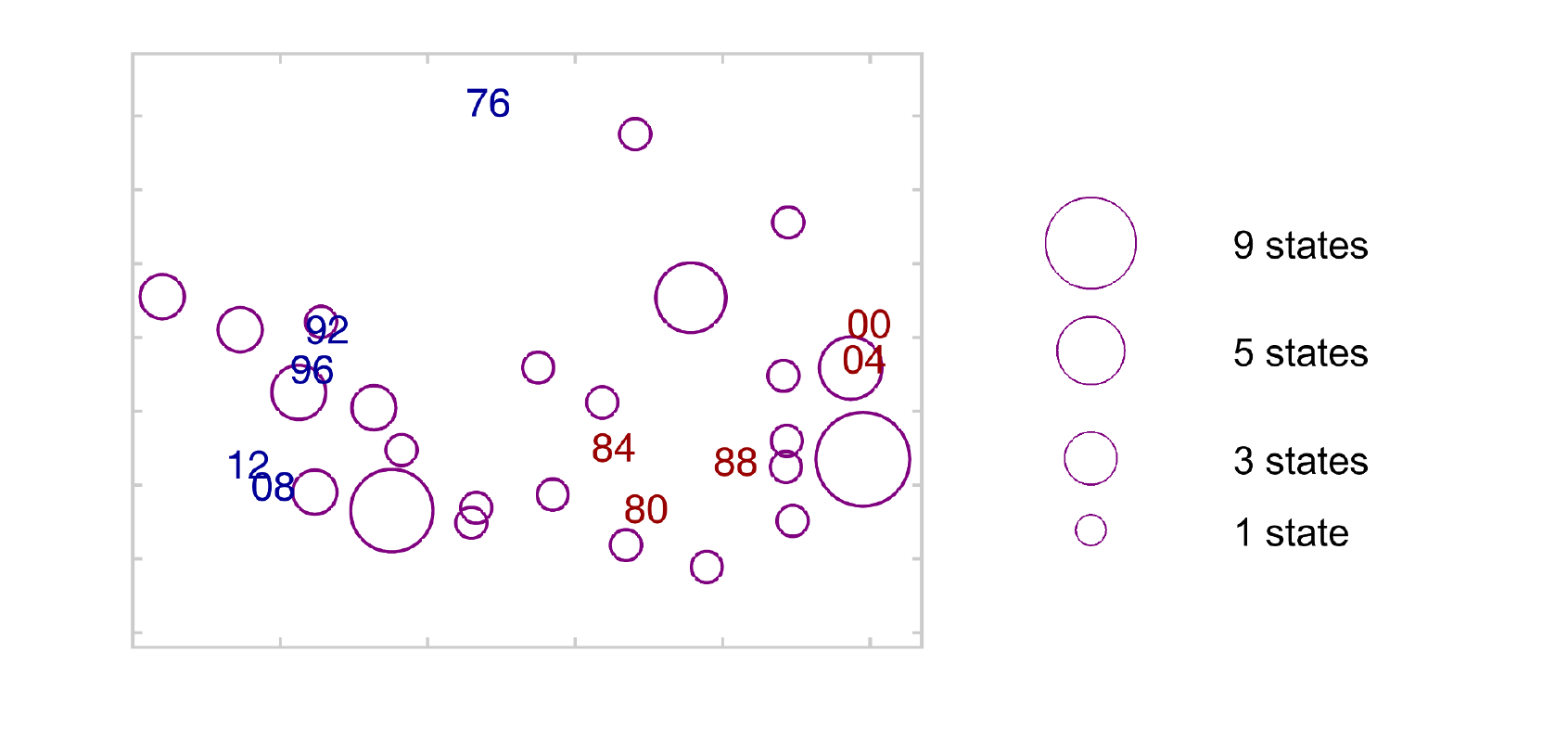}
\caption{{\bf BiFold bubble plot of US presidential election voting - by state (1976-2012).}  The BiFold layout based on state voting records, where $b_{ij}=1$ if state $i$ voted for the winning candidate in election $j.$  Bubble size (area) is proportional to the number of states at that position, with the smallest circle indicating one state.  
Each two-digit year is positioned at the BiFold coordinates of the elected president.     \label{Fig_State}}
\end{figure}

Consider the state-level votes for the United States presidential elections for the period from 1976 to 2012.
There are $51$ decision makers ($50$ states plus the District of Columbia) and a total of $10$ decisions, resulting in a data matrix $B=[b_{ij}]_{51 \times 10}$ where
\begin{equation}
b_{ij}=
\begin{cases}
1 & \text{if state $i$ voted for the winner of election $j$} \\
0 & \text{otherwise}. \\
\end{cases}
\end{equation}
As with the Southern Women example, we seek a  low-dimensional visualization which captures the within-class relationships (state-to-state and president-to-president) while also accounting for the between-class relationships (state-to-president).
To quantify these relationships, we define the dissimilarity between two states as the fraction of elections for which they voted differently; the dissimilarity between two elections is computed as the fraction of states which voted differently in those elections; finally, the dissimilarity between state $i$ and election $j$ is quantified as $1-b_{ij}$. Given these dissimilarities, Fig.~\ref{Fig_State} visualizes these election results, where coordinates are determined using BiFold.

\begin{figure}[htbp]
\centering
\includegraphics[width=3.5in]{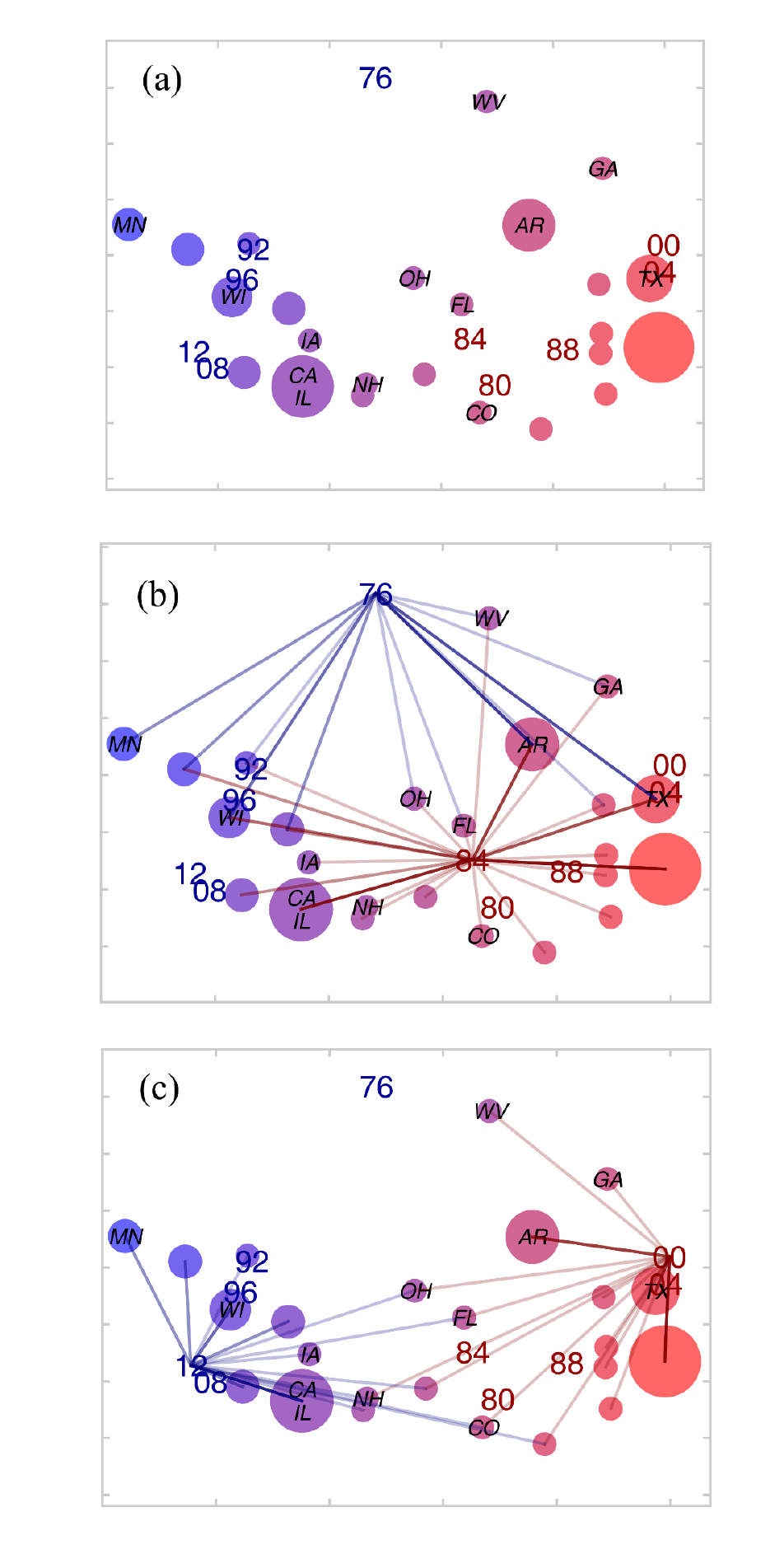}
\caption{{\bf BiFold of US presidential election - visual enhancement.} (a) States are colored based on the fraction of elections for which they supported the Republican (red) or Democrat (blue) presidential candidate.  (b) \textcolor{black}{Lines are added to connect Carter (`76) and Reagan (`84) to their respective ``supporting" states in those elections.} The boldness of the lines are proportional to the number of supporting states.  (c) Similar support representation for Bush `00 and Obama `12 illustrate the strong partisan nature of those elections.   \label{Fig_State2}   }
\end{figure}

Using BiFold for positional layout, we may encode additional information using other aesthetics.  In Fig.~\ref{Fig_State2}, states are colored according to party affinity (based on the fraction of these 10 elections in which that stated voted for the presidential candidate from that party, with Republican in red and Democrat in blue).

Aided by the additional encoded information, the BiFold layout yields some interesting observations.
\begin{itemize}\vspace{-0.05in}
\item Not surprisingly, the primary coordinate axis (left/right) strongly encodes the party affinity (blue state/red state).
\item Over time, the {\it election} positions have (generally) moved toward the left/right extremities, capturing the increasing partisanship of the elections.
\item Most of the purple colored  ``swing states'' are near the center of the visualization, which implies that they align with most of the election winners, with slight variation based on the particular set of presidents that they supported.  As interesting exception, West Virginia lies far above the main cloud, which we attribute to its trend of having often supported the non-winning candidates.
\item Noting several ``paired'' election coordinates, we observe that such pairs associate with two-term presidents, likely because their constituent support did not change much between elections.
\item  Positional outlier Carter `76 reflects support from a non-typical coalition of states, likely attributed to Carter being the first president elected from the Deep South \textcolor{black}{since} the Civil War.  Reagan `84 is the most centrally positioned, reflecting broad national support.  Figure~\ref{Fig_State2}b connects each of these elections to the supporting states.
\item  Comparing Bush `00 to Obama `12 (Fig.~\ref{Fig_State2}c) we see both elections driven primarily by partisan support.
\end{itemize}

We remark that any of the visually indicated hypothesis should be viewed as {\it exploratory} and be confirmed by additional quantitative analysis (as would also be appropriate for most other data visualizations).  However, we note that the BiFold visualization motivates a rich palette of such hypotheses, many of which directly exploit the between-class information.

\subsubsection{Senate congressional roll call votes.}
We consider the  voting record of the United States (U.S.) Senate. The U.S. legislative body is composed of two chambers, known as the Senate and the House of Representatives.  A particular congress serves for two years with this time frame divided into two sessions.   
We focus on voting data from the 112th congress, first session of the Senate, which conducted 235 roll call votes. 
For each roll call vote, there are at most 100 senators.  However, the replacement of Senator Ensign by Senator Heller mid-session leads to a data matrix with 101 rows and 235 columns,  recording the action of senator $i$ on bill $j$ 
\[
b_{ij}=
\begin{cases}
1 & \text{if a ``yes'' vote} \\
0 & \text{if a ``no'' vote},\\
\end{cases}
\]
If senator $i$ did not act on bill $j,$  entry $b_{ij}$ is undefined and is treated as {\it missing} data.  (See {\it Materials and Methods} for treatment of missing data.)

As with the previous examples, the goal (achieved by BiFold) is to obtain an embedding that captures both the within-class relationships (senator-to-senator and bill-to-bill) and the between-class relationships (senator-to-bill). From the data, we quantify the dissimilarity between two senators (bills) as the estimated probability that they vote (were voted) differently.  Dissimilarity between senator $i$ and bill $j$ is the estimated likelihood that senator $i$ objects to bill $j.$

\begin{figure}[htbp]
\centering
\includegraphics[width=6in]{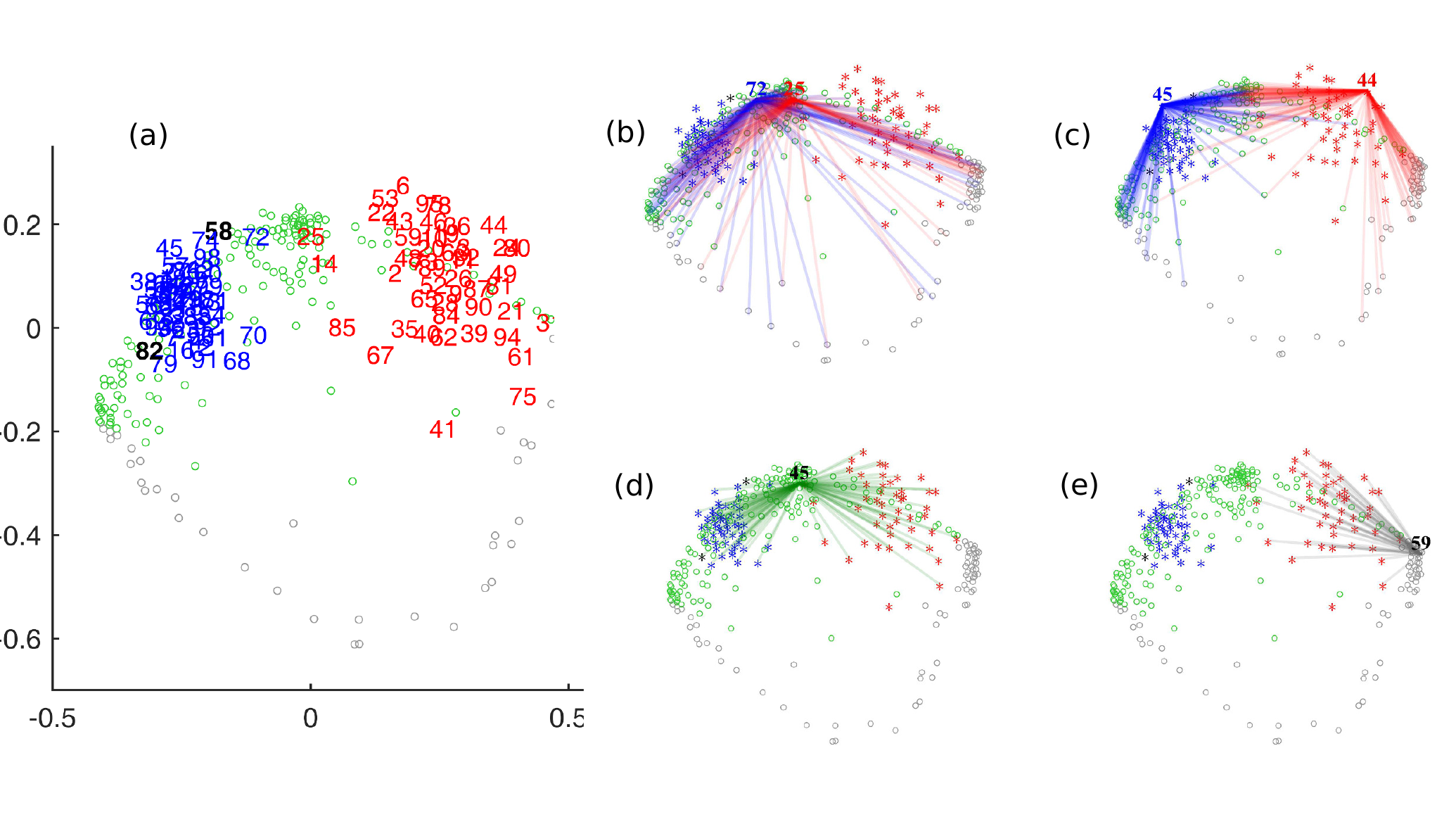}
\caption{
{\bf BiFold of voting records: US Senate, 112th Congress, Session1.}  (a) BiFold layout of Senate voting data.  Senators are numerically labeled (based on alphabetical order), with Republicans in red, Democrats in blue, and Independents in black.  Votes that passed are colored green, while those that did not pass are gray.  Observe that the two ``Independents,''  Senators Lieberman (58) and Sanders (82) align with the broad democratic party cluster, but both near its fringe.  Republican Senators Snowe (85),  Brown (14), and Collins (25) appear (by the BiFold plot) to have the most liberal voting record of their party.
 (b-d) Further aspects of the BiFold Senate layout. (b) ``Nearby'' senators Collins (R-25) and Nelson (D-72),  with lines connecting each senator position to their respective ``yes'' votes.  Both show strong bipartisanship in their voting records. 
 (c) Senators Inhofee (R-44) and Inouye (D-45) have very few votes that ``reach across the aisle.''
(d) Bills positioned near the center of the plot have broad, bipartisan support.  Vote 45 unanimously confirmed Amy Jackson as a US District judge. The lines connect this vote to those senators who voted in support.  (e) Vote 59, supported only by the Republican caucus, failed to pass.  It is situated ``far'' from the cloud of democrats, but near the Republicans.  This resolution would restrict use of Department of Defense Funds to carry out provisions of the Patient Protection and Affordable Care Act. \label{FigSenateDetails}}
\end{figure}

As shown in Fig.~\ref{FigSenateDetails}, the BiFold plot clearly shows the two-party structure of the senate, allowing for convenient visual comparison of the relative ``spread" of the parties, and identification of senators that are ``moderate" versus those that are more ``extreme" (Fig.~\ref{FigSenateDetails}, top panels).
The pattern of bills revealed by BiFold is reminiscent of the diamond structure previously identified from classical MDS~(\cite{Porter2005}).  In addition, BiFold provides visual information regarding the relationships between bills and senators by positioning bills ``close to'' the senators supporting them.  This unique feature enables a clear classification of the main clusters of bills as shown in  Fig.~\ref{FigSenateDetails}: 
\begin{itemize}
\item Bills in the ``left'' (liberal) cluster received strong support from the Democratic Senators;
\item Bills in the ``right'' (conservative) cluster received strong support from the Republicans;
\item Bills in the ``top" (bipartisan supportive) cluster were strongly supported by both parties, as visually being ``pulled" between the two parties;
\item Bills in the ``bottom" (bipartisan opposition) cluster are pushed far away from both parties, indicating bills that were supported only by a small number of senators.
\end{itemize}
Thus, by simultaneous embedding of both the senators and the bills, the BiFold visualization not only captures patterns within the senators and those within the bills, but also reveals salient features of the senator-bill cross relations.

\subsection{Association Datasets: A Recipe --- ingredient dataset}
We envision the BiFold approach to be broadly useful, certainly beyond the visualization of voting data.
Another important category of binary data captures the association between ``members" and ``affiliations." A key feature of such association datasets is that the non-association relations carry little information compared to the association relations; in sharp contrast, in a voting dataset the ``yes" and ``no" votes both convey valuable information about the relation between the decision makers and the choices.   
Association datasets are often collected to form sparse, bipartite networks, where sparsity arises from the reality that there are (typically) many more non-associations than associations in these data.

We focus here on a specific example relating {\it recipes} with their included {\it ingredients}.
A recipe defines a procedure for cooking, along with a list of food items (ingredients) used in the recipe.  Gathering this data over a broad spectrum of recipes allows us to more completely understand how ingredients are used in combination, which may vary from one cuisine to another.  As our data source, we consider the recipe-ingredient association dataset generated in \cite{Ahn2011}, which assembled over $50,000$ recipes taken from two American and one Korean online repository. The data is (again) represented by a  matrix $B,$ where $b_{ij}=1$ indicates that recipe $i$ contains ingredient $j,$ and $0$ otherwise. To proceed with the BiFold approach, we must define the dissimilarities between the entities: Recall that in the voting examples, both a ``1'' (a yes vote) and a ``0'' (a no vote) contain actual information regarding a voter's opinion. 
In contrast, in the recipe-ingredient dataset, a given recipe typically includes only a small fraction of all available ingredients and  carries essentially no information on those ingredients that are not  used in the recipe.

Between-class dissimilarity measure is as before, $\delta_{ij}=1-b_{ij}.$  However, the within-class dissimilarities require more careful consideration. If we were to quantify the dissimilarity between two recipes in the same way that we did for two voters, we would conclude that most recipes are very ``similar.''  This apparent similarity is artificial, resulting not from commonality of ingredients they share, but due to the overwhelmingly large set of ingredients that neither recipe contains. A dissimilarity measure that symmetrically incorporates ``1"s and ``0"s will therefore be dominated by the sparsity of the data rather than the actual relation between the entities of interest.
In this context, we would consider the 0s as carrying relatively little information.  As such, the {\it Jaccard distance} provides a natural measure of dissimilarity~\cite{Levandowsky}, where we treat rows (or columns) of $B$  as a characteristic function indicating set membership. For two recipes, the Jaccard distance is
\begin{equation}\vspace{-0.05in}
\label{eqJaccard}
J^{\mbox{\footnotesize{R}}}=1-\frac{\mbox{\#  ingredients shared by the two recipes }}{\mbox{\# ingredients needed to make both recipes}}.
\end{equation} 
Likewise, the Jaccard distance between two ingredients is
\begin{equation}
J^{\mbox{\footnotesize{I}}}=1-\frac{\mbox{\#  recipes using both ingredients }}{\mbox{\# recipes using either ingredient}}.
\end{equation} 

\begin{figure}[htbp]
\centering
\includegraphics[width=5.2in]{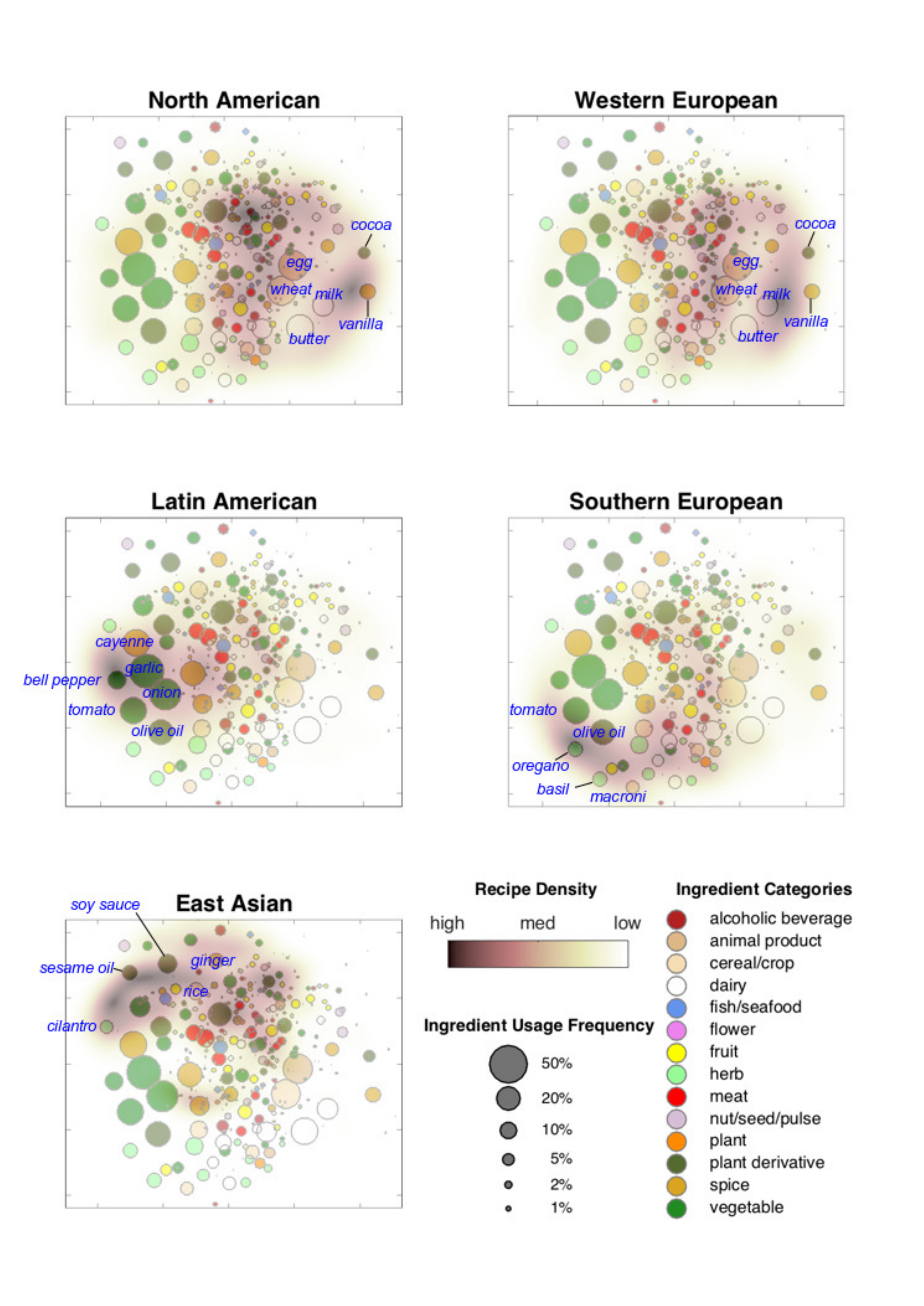}
\caption{{\bf BiFold visualization of recipes and ingredients.} Ingredients are plotted as circles with area proportional to the frequency of usage in recipes and coloring indicating its category. 
For each of the five cuisines (North American, Western European, Latin American, Southern European, and East Asian) we plot a density plot of the recipes from that cuisine along with the full set of ingredients. 
The BiFold layout enables visual exploration of the joint recipe-ingredient space across cuisines. For example, the central cloud  contains protein groups (primarily meat) shared by all cuisines. Each cuisine is visually associated with a few ``signature" ingredients, with examples include {\it basil} and {\it oregano} (Southern European), {\it sesame oil} and {\it soy sauce} (East Asian), and so on. The BiFold plots allow us to ``see'' that certain cuisines are likely similar (North American --- Western European, and Latin American --- Southern European).  \label{Fig_Recipe}}
\end{figure}

In addition to the recipe-ingredient relationship information, the original dataset also categorized each recipe as belonging to a particular {\it cuisine}.  We focus our analysis on a random subsample (of 1000 recipes) of the five cuisines in the original dataset that contain more than 1000 recipes.  We compute a 2-D BiFold embedding to support visualization of this reduced dataset.   In Fig.~\ref{Fig_Recipe}, we use the BiFold coordinates to plot food ingredients (circles, colored by ingredient category), with that layout the same for all five cuisines.  
Each cuisine is visualized in its own panel, where we use a density plot  to capture the distribution of recipes from that cuisine.

As expected, ingredients that are commonly used together in recipes are positioned near each other in the plot, and recipes with similar ingredients appear close together as well.  
A unique outcome of applying BiFold to this data is that we may now visually associate {\it ingredients} to cuisines, whereas the original data only associates {\it recipes} to cuisines,  facilitating an entirely new level of interpretation enabled by embedding  both recipes and ingredients using a common coordinate frame:
\begin{itemize}
\item From the collection of cuisine plots, we can visually identify similar cuisines (North America --- Western Europe, Latin America --- Southern Europe).
\item The East Asian cuisine appears visually distinct from the western heritage cuisines.
\item The protein group, primarily meat, appear centrally in the figure of ingredients, with all the cuisines showing significant density in that region of the plot.  (In other words, the meat group does not identify any particular cuisine.)
 \item The density plots allow to visually identify certain ingredients as the ``signature'' of a cuisine: {\it basil} and {\it oregano} (Southern European); {\it sesame oil} and {\it soy sauce} (East Asian) ;  {\it cocoa} and {\it vanilla} (North American and Western European).
\end{itemize}

\section{Discussion}
\begin{figure}[htbp]
\centering
\includegraphics[width=4in]{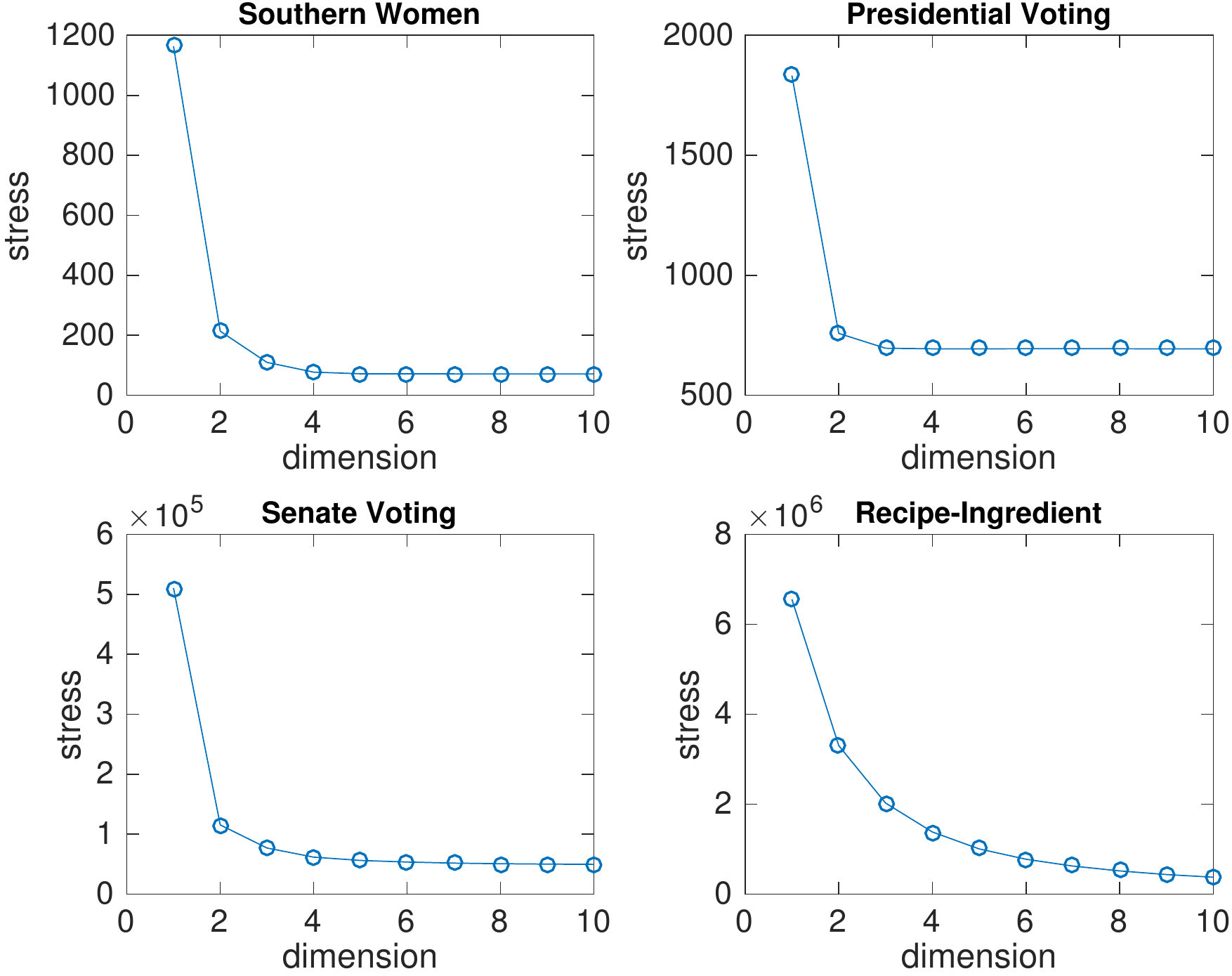}
\caption{{\bf Minimal stress vs. dimension in BiFold.} For each of the example dataset used in this paper, we plot the minimal BiFold stress obtained by the SMACOF algorithm (see {\it Materials and Methods} for details) as a function of the embedding dimension.
\label{Fig:BiStress}}
\end{figure}

The BiFold framework described in this article has primarily focused on a fixed, binary dataset, interpretable as associations between two types of objects.  We consider that framework to be broadly applicable to  datasets describing relationships between entities from different classes, where we want to be able to {\it simultaneously visualize the different classes} such that {\it visual distance} can be associated to a {\it dissimilarity measure}, both within class and between classes. 
For the datasets examined, we would remark that although the {\it knowledge discovery} facilitated by the visualization are possibly achievable by other analysis techniques, BiFold has a unique ability to simultaneously visualize those discoveries.
Note that the extent to which BiFold plot (or any visualization) reflects the actual similarities and dissimilarities between objects in the dataset---as measured by the stress function---depends intrinsically on the dataset itself. 
In typical real-world datasets, the representation would not be perfect, even if the dimensionality of embedding is large. For the datasets considered here, we find that in the Southern Women example, as well as the two US voting examples, a low-dimensional (2-D or 3-D) BiFold embedding achieves an almost minimal stress which cannot be further decreased by increasing dimensionality (see Fig.~\ref{Fig:BiStress}), supporting the notion that the opinions are well expressed by a low-dimensional model. 
On the other hand, for the recipe-ingredient example, increase of dimensionality beyond 3-D continue to decrease stress and improve the match to the original data (Fig.~\ref{Fig:BiStress}), suggesting an enormous diversity and complexity in the cuisine space which cannot be accounted for using just a few variables or parameters.

In addition, we note that the BiFold framework described here may be easily extended to a number of interesting and related problems:
\begin{itemize}
\item  As an (almost trivial) extension, we note that interpretation of the data as representing a bipartite network implies that BiFold could act as a graph layout algorithm for bipartite network data.
\item  BiFold can be viewed as a generalization of several other classical techniques which can be recovered by specific choice of parameters:
\begin{itemize}
\item $w^{(xy)}_{ij}=0$: Only within-class dissimilarities are considered, yielding separate MDS embeddings of the two types of objects~\cite{CoxBook,BorgBook}.
\item $w^{(x)}_{ij}=w^{(y)}_{ij}=0$: only between-class dissimilarities are considered, yielding  an {\it unfolding} of the data~\cite{CoxBook,BorgBook,Bennett1960}.
\end{itemize} 
\item  The entries in the data matrix, $B,$ need not be binary, but could represent a continuous or ordinal variable, such as ratings, rankings, or preferences.
\item  Some dataset might naturally contain more than two groups, such as {\it actors, movies, and viewers.}  Such datasets can be treated as multipartite, rather than bipartite data.  We envision a natural extension of BiFold, where the joint dissimilarity and weighting matrices must be appropriately constructed based on the within-group and between-group relationships. 
\item  We focused on Hamming distance and Jaccard distance to compute within-in class dissimilarity, with each providing a natural interpretation for the datasets considered.  We note that the BiFold framework is not dependent upon any particular choice of dissimilarity measure, and a reasonable practitioner may choose other methods for defining dissimilarities (and weights) that might be appropriate for their data.  The BiFold approach - based on the joint dissimilarity matrix, will still provide a means to develop the joint visualization.
\item  For some of the methods, we interpret the raw (binary) from Bayesian perspective, but with uninformed prior.  That approach could easily added to accommodate other {\it a priori} understanding of the data.
\item  For dynamic datasets (parameterized by time, for example) each data ``snapshot'' would yield a BiFold layout.  A stress functional that incorporates a regularity condition in time could compute an optimal sequence of layouts, computed over many snapshots.
\end{itemize}

As caution, we note some of the challenges associated with analysis via the BiFold framework:
\begin{itemize}
\item Computational complexity of the stress minimization as an optimization problem using the SMACOF algorithm is roughly $O(n^4)$ for reaching at a local,  approximate solution.  As such the current implementation of stress minimization will likely struggle with very large datasets.  Because the technique is meant to support visual knowledge discovery (human interaction), speed of visualization is important.  Data aggregation might be a way to handle large datasets, but the aggregation procedures will almost certainly be domain specific. 
\item  Comparing one BiFold layout to another (exploring parameter space) can be challenging in that the solution layout is rotation and reflection invariant.  Normalizing the orientation of the generated solution is important.  As additional complication, the configuration {\it solution} to the optimization problem is a local minimizer, so that solution may ``jump'' to a different minimizer under small changes in the data.  
\item The non-euclidean nature of the dissimilarity measures results in a dissimilarity matrix that is not necessarily well approximated by a low dimensional embedding.  Under such case, visually interesting effects may sometimes be an artifact of the data, particularly with sparse datasets. 
\end{itemize}
Despite these challenges, we note that the proposed BiFold framework developed here appears to have broad applicability in many settings related to complex networks, social sciences, and those areas of data analysis that focus on binary relations.

\section*{Materials and Methods}
\subsection*{\bf Datasets.} 
\begin{itemize}
\item The {\bf Southern Women dataset} is a popular dataset used in social network analysis. The dataset first appeared in the book ``Deep South: A Social Anthropological Study of Caste and Class"~\cite{SWdata} (pp. 148), and can also be found in several online network data repositories.
Collected in the 1930s in a small southern town Natchez (Mississippi, United States), the data records the participation of 18 women in a series of 14 informal social events over a nine-month period. Only the events for which at least two women participated are included in the dataset. Figure~\ref{Fig_SW} shows the data table without including the names of the women or dates of the events. We represent the dataset by a woman-by-event matrix $B=[b_{ij}]_{18\times14}$, where $b_{ij}=1$ indicates that woman $i$ attended event $j$, and $b_{ij}=0$ otherwise.
\item The {\bf U.S presidential election dataset} considered in this paper includes the state-level voting results of the United States presidential elections for the period from 1976 to 2012. The dataset, available at the U.S government archive (http://www.archives.gov/federal-register/electoral-college/), includes the state voting outcome from the $51$ voting entities ($50$ states plus the District of Columbia) for the past $10$ presidential elections. 
We alphabetically numbering the states from $1$ to $51$ by name, and the elections from $1$ to $10$ in chronological order.
We then represent the dataset by a state-by-president matrix $B=[b_{ij}]_{51\times10}$, where $b_{ij}=1$ indicates that state $i$ voted for the elected president in the $j$-th election, and $b_{ij}=0$ otherwise. For example, in all past $10$ elections Ohio has always voted for the president candidate who eventually won the election regardless of his party affiliation. Florida and Nevada both ``missed" one election: in the 1992 election, Florida voted for G.~H.~W.~Bush (the elected president was B.~Clinton); in the 1976 election, Nevada voted for G.~Ford (the elected president was J.~Carter). All three are well-known examples of ``swing" states characterized by flexible voting patterns and importance in determining the election outcome.
\item The {\bf U.S Senate Congressional Voting dataset} used in this paper is obtained from the congressional voting records of the 112th United States congress, first session of the Senate.  There are at most 100 senators at any time, with occasional need to replace a senator in mid session, which happened once during the voting portion of this session. 
As such the roll calls indicate 101 senators voting,  51 Democrats (D), 48 Republicans (R), and 2 Independents (I).   There were 235 recorded roll call votes, 167 passed and 68 rejected. 
We number the senators from 1 to 101 by last name, and the bills from 1 to 235 in chronological order.
We formulate data matrix $B=[b_{ij}]_{101\times235}$ by defining $b_{ij}$ using the voting of senator $i$ on bill $j$: for a ``yes" vote $b_{ij}=1$, for a ``no" vote $b_{ij}=0$. The abstained votes are treated as ``missing" data in the matrix (see the ``Treatment of partial and missing data" section below for details).
\item The {\bf recipe-ingredient dataset} is retrieved from the Supplementary Information of Ref.~\cite{Ahn2011}, a paper that studied the similarity and difference in food pairings across different geographical regions. The dataset contains more than 50,000 recipes extracted from three cuisine websites: allrecipes.com, epicurious.com, and menupan.com. The recipes were divided into 11 geographical regions, covering $\sim$50 popular cuisines around the world.
The recipes and ingredients are indexed.
Focusing on the 5 geographical regions (cuisines) that contain over 1000 recipes, 
we construct data matrix $B=[b_{ij}]_{5000\times 335}$, with $b_{ij}=1$ if recipe $i$ contains ingredient $j$.  This subsample of the original dataset contains 1000 randomly selected recipes from each of the 5 selected cuisines: East Asian, Latin American, North American, Southern European, and  Western European. The subsampled data contains a total of 335 different ingredients.
\end{itemize}

\subsection*{\bf The BiFold framework: dissimilarity measures, weights, and stress minimization.}
The BiFold framework describes a general approach to produce a low-dimensional embedding from a data matrix, where that matrix encodes the relationship between two classes of objects. 
First, one needs to create a {\it joint dissimilarity matrix} using some appropriate within-class and cross-class dissimilarity measures as well as scaling to make the within-class and cross-class dissimilarities commensurate. Secondly, one needs to construct a {\it weighting matrix} to reflect the relative focus to be given to the computed dissimilarities. Finally, the BiFold embedding is obtained by minimizing a weighed \textit{stress function} similar to the determination of an MDS solution.

We now present the mathematical details of the BiFold procedure. For a given data matrix $B=[b_{ij}]_{m\times n}$, a $d$-dimensional BiFold embedding is based upon minimization of the multivariate stress function
$S:\mathbb{R}^{d\times(m+n)}\rightarrow\mathbb{R}$, defined as
\begin{equation} \label{eq:stress}
S(\bm{z}_1,\bm{z}_2,\dots,\bm{z}_{m+n})=\sum_{k,\ell=1}^{m+n}  w_{k\ell} \Phi(\|\bm{z}_k-\bm{z}_\ell\|_2,\delta_{k\ell}).
\end{equation}

\begin{itemize}
\item The joint dissimilarity matrix is given by
\begin{equation}
\Delta = \begin{bmatrix}
      \alpha_x\Delta^{(x)} & \alpha_{xy}\Delta^{(xy)}+\beta\bm{1}  \\[0.2em]
      \alpha_{xy}\Delta^{(yx)}+\beta\bm{1} &   \alpha_y\Delta^{(y)} \\
     \end{bmatrix},
\end{equation}
where $\Delta^{(x)}=[\delta^{(x)}_{ij}]_{m\times m}$ and $\Delta^{(y)}=[\delta^{(y)}_{ij}]_{n\times n}$ are the within-class dissimilarity matrices and $\Delta^{(xy)}=[\delta^{(xy)}_{ij}]_{m\times n}$ is the cross-class dissimilarity matrix ($\Delta^{(yx)}=\Delta^{(xy)\top}$).  The parameters: $\alpha_x$, $\alpha_y$, and $\alpha_{xy}$ provide flexible scaling of the within-class and cross-class distances in the embedded space, while $\beta$ can be used to visually translate the type-1 objects away from the type-2 in the embedding.

\item The weighting matrix is defined as
\begin{equation}\label{eq:weightmatrix}
W = \begin{bmatrix}
     W^{(x)} & W^{(xy)}  \\[0.2em]
      W^{(yx)} &  W^{(y)} \\
     \end{bmatrix},
\end{equation}
where $W^{(x)}=[w^{(x)}]_{m\times m}$, $W^{(y)}=[w^{(y)}]_{n\times n}$ are the within-class weighting matrices and $W^{(xy)}=[w^{(xy)}]_{m\times n}$ is the cross-class weighting matrix ($W^{(yx)}=W^{(xy)\top}$). 
\item As typical choice for the above stress function $S$ is to let $\Phi(d,\delta)=(d-\delta)^2.$  For a given dissimilarity and weight matrix, this fully specified stress function may then be minimized to obtain coordinates $\{z_1, \ldots,z_{m+n}\}.$  
\end{itemize}

\subsection*{\bf Dissimilarity measures and weights used in the examples.}
In the data matrix $B=[b_{ij}]_{m\times n}$ of the {\it Southern Women dataset}, $b_{ij}=1$ if woman $i$ attended event $j$ and $b_{ij}=0$ otherwise. For the BiFold plot in Fig.~\ref{Fig_SW}, we used the following within-class and cross-class dissimilarities:
\begin{equation}
\begin{cases}
\mbox{\small (woman-to-woman dissimilarity)}&\delta^{(x)}_{ij}=\sum_{k=1}^{n}|b_{ik}-b_{jk}|,\\
\mbox{\small(event-to-event dissimilarity)}&\delta^{(y)}_{ij}=\sum_{k=1}^{m}|b_{ki}-b_{kj}|,\\
\mbox{\small(woman-to-event dissimilarity)}&\delta^{(xy)}_{ij}=1-b_{ij}.
\end{cases}
\end{equation}
Then, to balance the spread of the points from the two classes in the embedding, we set the scaling parameters $\alpha_x=1/n$, $\alpha_y=1/m$, and $\alpha_{xy}=1$. The shifting parameter $\beta=0$. All entries of the joint weighting matrix $W$ equal to $1$. These choices were made primarily for simplicity and are unlikely to be appropriate for the other, much larger datasets considered in the paper. Below we develop a set of dissimilarity measures and corresponding weights suitable for two common types of data matrices encoding voting and association relations, respectively.

\begin{itemize}
\item {\bf Voting data: the BiFold Bernoulli Method.} Where the data matrix $B$ represents `voting' data, such that $b_{ij}$ indicates that object $X^i$ voted positively for object $Y^j,$ one may consider that the preference selection (`1' or `0') is a forced binary decision on a continuous variable that represents preference.  One model for this situation would be to view $b_{ij}$ as the observation of the {\it forced decision outcome}, treated as a Bernoulli trial, where Bernoulli parameter 
$p:=p_{ij}=:p_{ij}^{(xy)}$ is not known. 
(For real data sets of voting data, we treat `yes' as `1' and `no' as `0.'  As a third outcome, sometimes a voter will `abstain' on a particular vote, which we view as ``missing data'' with technique described below.)
Applying this model within a group (for example, within group 1) we could assert a Bernoulli process with
$p:=p_{ij}^{(xx)}$
the (unknown) probability that object $X^i$ and $X^j$ would vote the same way on an arbitrarily selected vote.  Comparing rows $i$ and $j$ in the data matrix $B$ would provide $n$ observations of outcomes from that Bernoulli process.  Comparison of columns treated in the same way, would represent $m$ observation of the Bernoulli process associated to objects $Y^i$ and $Y^j.$ Ideally, we would like to construct a BiFold configuration using dissimilarities
    computed from the {\it actual} values for preference --- the unknown values for $p_{ij}^{kl}.$   Instead, we must assign dissimalities from estimated probabilities , $\delta_{ij}^{(*)}:=1-\hat{p}_{ij}^{(*)}.$  
Following standard development for estimating proportions, we count the number of {\it within group differences}  between pairs of entities in each class :
\begin{align}
s_{ij}^{(x)}&= \sum_k |b_{ik}-b_{jk}|,\\
s_{ij}^{(y)}&= \sum_k |b_{ki}-b_{kj}|.
\end{align}
For the cross-class data, we pool all observations to define an average rate of positive voting:
\begin{equation}
\bar{p}=\frac{\sum\limits_{i,j} b_{ij}}{nm}.
\end{equation}

Because we have significantly more observations for the `within class' data, we expect those estimates to be more accurate.  Consequently, we choose weights $w_{ij}$ proportional to the information content.  Borrowing from approaches used in regression of heteroscedastic data, we weight the error term (stress) inversely as the (estimated) variance in the observation, as applied in equations \eqref{eq:stress} and \eqref{eq:weightmatrix}.  We focus on three primary alternatives for the estimation of the parameters and the variance: (1) Bayesian, with uniform prior; (2) Bayesian, with Jeffreys' prior; and (3) Non-Bayesian, maximum likelihood estimate.  Table 1 shows the resultant formulas associated to these methods. We note that the specific Bayesian approaches described assume no prior belief regarding the parameters $p_{ij}.$  However, the concept is obviously easily generalized to those cases where prior information is available, where one would simply encode that knowledge into assumed prior distribution.

\item {\bf Association data: the BiFold Membership Method.}  For association data (such as the {\bf recipe-ingredient} dataset), the sparse biadjacency matrix $b_{ij}=1$ indicates an association between object  $i$ from class $x$ with object $j$ from class $y.$  
\textcolor{black}{Unlike the case of voting datasets a ``0" in an association dataset carries relatively little information as opposed to a ``1". This asymmetry, if not accounted for appropriately, will result in an embedding (and visualization) that is dominated by the count of 1s instead of revealing more useful features.
}

Between class dissimilarity measure is quantified as
\begin{equation}
\delta_{ij}^{(xy)}=1-b_{ij}.
\end{equation}
The within-class dissimilarities are computed using a Jaccard distance.  Specifically, for two objects represented by rows $i$ and $j$  of the matrix $B$, their dissimilarity is given by 
\begin{equation}
\delta^{(x)}_{ij}=1-\frac{\sum_k b_{ik}b_{jk}}{\sum_k \left(b_{ik}+b_{jk}-b_{ik}b_{jk}\right)}.
\end{equation} 
Likewise, the dissimilarity between columns $i$ and $j$ is computed as
\begin{equation}
\delta^{(y)}_{ij}=1-\frac{\sum_k b_{ki}b_{kj}}{\sum_k \left(b_{ki}+b_{kj}-b_{ki}b_{kj}\right)}.
\end{equation} 
\end{itemize}

For weights, we treat $b_{ij}=1$ as representing unit information, while $b_{ij}=0$ carries no information, so that 
\begin{equation}
w_{ij}^{(xy)}=1-b_{ij}.
\end{equation}
For within class, the weights are computed by counting the number of common ``1's,'' yielding
\begin{equation}
w^{(x)}_{ij}=\sum_k b_{ik}b_{jk}, \qquad   w^{(y)}_{ij}=\sum_k b_{ki}b_{kj}.
\end{equation} 
As a result of the typical sparsity in such dataset, matrix $W$ will also be sparse.  We remark that
\begin{equation}
w_{ij}=0 \iff \delta_{ij}=1,
\end{equation}
meaning that under this condition of maximal dissimilarity of $i$ with $j,$ that particular dissimilarity does not directly affect the computed stress functional or the resultant BiFold embedding.  Without this weighting scheme, a sparse association dataset would be completely dominated (visually) by the large number of objects forced to lie at the outside of the unit ball because most objects are `very far' from most other objects.

\smallskip\noindent{\bf 4. Stress minimization.}
After formulating a stress function~\eqref{eq:stress} and embedding dimension $d$, a BiFold representation of the data is obtained by minimizing the stress function over the coordinates of $m+n$ points in a $d$-dimensional Euclidean space. This optimization problem is within the class of MDS problems, with several alternative tools available to find a local minimum~\cite{CoxBook,BorgBook}. For the BiFold plots reported in this paper, the stress minimization is done via the (iterative) SMACOF algorithm~\cite{BorgBook}. For reproducibility of results, for the initial iteration of the algorithm, the starting configuration for the coordinates is obtained by a classical MDS solution of the joint dissimilarity matrix (without weighting).  After applying the SMACOF algorithm to obtain a set of coordinates, we further perform a PCA (principal component analysis) to standardize the alignment, noting that the stress function is invariant under such transformations. As a consequence, in all BiFold plots the horizontal axis is the principal direction.

\smallskip\noindent{\bf 5. Treatment of partial and missing data.}  For real datasets, the choice of methods for dealing with missing data can become a critical component of the data processing.  In general, the BiFold approach admits a very reasoned approach that does not depend upon imputation and remains robust in a wide variety of datasets.  The key enabler is recognizing that data matrix $B$ contains $mn$ pieces of information, while the solution (a configuration) allows just $d \times (m+n)$ free variables.  Under typical scenarios, with the visualization dimension $d=2$ or $d=3,$ and $n,m\gg d,$ we may view this as the data matrix as providing significant amount of ``redundant'' information.  In the same way that a regression line should not suffer too much if a small fraction of the data set is removed, a similar robustness should persist in the BiFold visualization.  As such, we follow two general guidelines when dealing with missing data:
\begin{enumerate}
\item Use only available data when computing dissimilarities $\delta_{ij}.$
\item Weights $w_{ij}$ should be selected to account for the {\it actual} (non missing) data that is used to compute the associated dissimilarity.
\end{enumerate}

Consider, for example, the congressional voting data described above.  For these data, it is typical that not all senators would vote on every bill.  Some may ``abstain'' during the roll call, but others may simply not be present.  In this case, a typical dataset structure might assign 
\begin{equation}
b_{ij} = \mbox{NA}
\end{equation}•
if senator $i$ did not vote on bill $j.$ To perform BiFold under this condition of missing data, we proceed as follows:
\begin{itemize}
\item If $b_{ij} = \mbox{NA}$ then $w_{ij}^{(xy)}=0,$ and $\delta_{ij}^{(xy)}=c,$ where $c$ is an arbitrary, finite constant.
\item For within group differences for group 1, define index sets $\kappa_{ij}$ as
\[
\kappa_{ij} = \{k| b_{ik} \neq  \mbox{NA}, b_{jk}  \neq  \mbox{NA} \},
\]
compute
\begin{equation}
s_{ij}^{(x)}= \sum_{k \in \kappa_{ij}}  |b_{ik}-b_{jk}|,
\end{equation}
and determine the number of information elements as
\begin{equation}
n_{ij}^{(x)}= |\kappa_{ij}|.
\end{equation}
\item Apply Table 1 formulae to compute $\delta_{ij}^{(x)}$ and $w_{ij}^{(x)}$, replacing $n$ by $n_{ij}$.
\item Use similarly modified formulas to compute $\delta_{ij}^{(y)}$ and $w_{ij}^{(y)}.$
\end{itemize}

After forming the data matrices $\Delta$ and $W,$ then we may simply minimize the weighted stress to determine an coordinate representation.

\begin{table*}[h]
\centering
{\bf Table 1: Bifold Bernoulli methods: coefficient estimation formulas for the distances}\vspace{0.1in}
\begin{tabular*}{\hsize}{@{\extracolsep{\fill}}ccccccc}
\hline 
\multirow{2}{*}{Groups} & \multicolumn{2}{c}{Uniform prior} & \multicolumn{2}{c}{Jeffreys' prior} &  \multicolumn{2}{c}{Non-Bayes}  \\ 
&$\delta_{ij}$ & $1/w_{ij}$ & $\delta_{ij}$ & $1/w_{ij}$ & $\delta_{ij}$ & $1/w_{ij}$ \\
\\
$1 \leftrightarrow 2$ & $\frac{2-b_{ij}}{3}$   & $\bar{p}(1-\bar{p})$
           &   $\frac{3/2-b_{ij}}{2}$  & $\bar{p}(1-\bar{p})$
           &    $1-b_{ij}$ &  $\bar{p}(1-\bar{p})$   \\

\rule{0pt}{4ex}  $1 \leftrightarrow 1$ & $\frac{s_{ij}^{(11)}+1}{n+2}$ &   $\frac{\delta_{ij} (1-\delta_{ij})}{n}$
           &  $\frac{s_{ij}^{(11)} +1/2}{n+1}$ &   $\frac{\delta_{ij} (1-\delta_{ij})}{n}$
           &  $\frac{s_{ij}^{(11)}}{n}$   &    $\frac{\left(s_{ij}^{(11)} +1/2\right)\left(n-s_{ij}^{(11)} +1/2\right)}{(n+1)^2 n}$       \\

\rule{0pt}{4ex}  $2 \leftrightarrow 2$ & $\frac{s_{ij}^{(22)}+1}{m+2}$ &    $\frac{\delta_{ij} (1-\delta_{ij})}{m}$
           &  $\frac{s_{ij}^{(22)} +1/2}{m+1}$ &    $\frac{\delta_{ij} (1-\delta_{ij})}{m}$
           &  $\frac{s_{ij}^{(22)}}{m}$ &    $\frac{\left(s_{ij}^{(11)} +1/2\right)\left(m-s_{ij}^{(11)} +1/2\right)}{(m+1)^2 m}$           \\
           \\
\hline
\end{tabular*}
\end{table*}

%

\begin{acknowledgements}
The authors wish to thank Daniel B. Larremore for useful feedback on the manuscript.
\end{acknowledgements}

\section*{Author contributions statement}
J.D.S. and J.S. designed the research. All authors contributed to methodological and algorithm developments, data collection, visualization and analysis. J.D.S and J.S. wrote the manuscript.

\textcolor{black}{\section*{Funding} This work was partially supported by a Clarkson University Provost Award, Army Research Office grants W911NF-12-1-0276 and W911NF-16-1-0081, and the Simons Foundation grant 318812. Any opinions, findings, and conclusions or recommendations expressed in this material are those of the author(s) and do not necessarily reflect those of the National Science Foundation.}

\section*{Competing interests} The authors declare no competing financial interests.



%
%

\newpage
\textcolor{black}{\section*{Figure Legends}}

\end{document}